\documentclass[aps,prb,twocolumn,superscriptaddress]{revtex4}
\usepackage{amsmath,amssymb,amsfonts,amsthm}
\usepackage[utf8]{inputenc}
\usepackage{graphicx}
\usepackage[caption=false]{subfig}
\usepackage[pdftex,bookmarks=false,colorlinks=true,linkcolor=blue,citecolor=blue,filecolor=black,urlcolor=blue]{hyperref}

\providecommand{\Z}{{\mathbb{Z}}}
\providecommand{\N}{{\mathbb{N}}}
\providecommand{\fv}{\mathbf{f}}
\providecommand{\rv}{\mathbf{r}}

\begin{document}

\title{Wigner localization in quantum dots from Kohn-Sham
density functional theory without symmetry breaking}

\author{Christian~B.~Mendl}
\affiliation{Mathematics Department,
Technische Universit\"at M\"unchen,
Boltzmannstra{\ss}e 3,
85748 Garching bei M\"unchen, Germany}

\author{Francesc Malet}
\affiliation{Department of Theoretical Chemistry and
Amsterdam Center for Multiscale Modeling, FEW,
Vrije Universiteit,
De Boelelaan 1083,
1081HV Amsterdam,
The Netherlands}

\author{Paola Gori-Giorgi}
\affiliation{Department of Theoretical Chemistry and
Amsterdam Center for Multiscale Modeling, FEW,
Vrije Universiteit,
De Boelelaan 1083,
1081HV Amsterdam,
The Netherlands}


\begin{abstract}
We address low-density two-dimensional circular quantum dots with spin-restricted Kohn-Sham density functional theory. By using an exchange-correlation functional that encodes the effects of the strongly-correlated regime (and that becomes exact in the limit of infinite correlation), we are able to reproduce characteristic phenomena such as the formation of ring structures in the electronic total density, preserving the fundamental circular symmetry of the system. The observation of this and other well-known effects in Wigner-localized quantum dots such as the flattening of the addition energy spectra, has until now only been within the scope of other, numerically more demanding theoretical approaches.
\end{abstract}

\maketitle

\section{Introduction}
\label{sec:intro}

The effects of strong electronic correlation in low-dimensional semiconductor nanostructures
have attracted large research interest for decades, both from purely fundamental and 
from applied points of 
view.\cite{GhoGucUmrUllBar-NP-06, GucGhoUmrBar-PRB-08, YanLan-RPP-07,
AusSteYacTseHalBalPfeWes-Sci-05, KriCreNilXuSamLinWacRei-PRB-11} 
The high degree of tunability of, e.g., quantum wires or quantum dots, nowadays easily realized 
in laboratories,\cite{AusSteYacTseHalBalPfeWes-Sci-05,KriCreNilXuSamLinWacRei-PRB-11} 
renders them a fertile playground to investigate strong-correlation phenomena.
For example, it is well known that, for sufficiently low densities, such finite systems
may display charge localization,\cite{CreHauJefSar-PRB-99,EggHauMakGra-PRL-99,
YanLan-PRL-99, ReiKosMan-PRB-00,FilBonLoz-PRL-01,PedEmpLip-PRB-02,GhoGucUmrUllBar-PRB-07} reminiscent 
of the Wigner crystallization of 
the bulk electron gas\cite{Wig-PR-34} and a consequence of the dominance of the Coulomb 
repulsion over the electronic kinetic energy. From the practical side, potential 
applications of Wigner-localized systems include the design and manipulation 
of qubits and quantum computing 
devices,\cite{WeiThoEgg-EPL-06, DesBoc-NP-08, TayCal-PRA-08, YanLan-RPP-07} or the realization 
of infrared sensors to control the electron filling in semiconductor 
nanostructures.\cite{BalEscMovPiPla-PRB-10}

Along with the fundamental and practical interest, strongly-correlated systems are 
well-known to pose serious challenges for the different theoretical approaches 
commonly used to study them. On the one hand, the configuration interaction (CI) 
method becomes numerically unaffordable if one wants to treat more than five or 
six electrons.\cite{ReiMan-RMP-02,RonCavBelGol-JCP-06,GhoGucUmrUllBar-PRB-07} 
By using coupled-cluster methods, which allow for a larger basis set, it has
very recently been shown that the number of particles can be raised up to 
twelve in two-dimensional quantum dots.\cite{WalWesLin-PRB-13}
Other wavefunction approaches, such as quantum Monte Carlo (QMC) 
methods\cite{GhoGucUmrUllBar-NP-06,GhoGucUmrUllBar-PRB-07,GucGhoUmrBar-PRB-08}
or density matrix renormalization group (DMRG),\cite{StoWagWhiBur-PRL-12} can
treat larger systems (still less than $\sim 10^2$ particles) but face
limitations as well if the correlations become too strong.\cite{GhoGucUmrUllBar-PRB-07} On the 
other hand, spin-unrestricted Hartree-Fock (HF)\cite{YanLan-PRL-99,YanLan-RPP-07} or 
density-functional \cite{BorTorKosManAbeRei-IJQC-05} approaches, much less computationally 
demanding, mimic the effects of strong correlation by breaking the spin and other symmetries 
of the system. This makes them much less reliable than wavefunction methods,
sometimes with unphysical results and controversial 
interpretations.\cite{AniZaaAnd-PRB-91,HarRasSaaPusNieNie-PRB-04,BorTorKosManAbeRei-IJQC-05,StoWagWhiBur-PRL-12}

Kohn-Sham (KS) density functional theory (DFT), in its original restricted 
formulation,\cite{HohKoh-PR-64,KohSha-PR-65} 
has been known for a long time to deliver very poor results when applied to 
strongly-correlated systems. 
The reason for this is not fundamental, as KS DFT is, in principle, an exact theory. The problem 
is that the available approximations for the exchange-correlation functional fail in the 
strongly-correlated regime, \cite{AniZaaAnd-PRB-91,CohMorYan-SCI-08,GhoGucUmrUllBar-PRB-07,AbePolXiaTos-EJPB-07} 
sometimes making it extremely difficult to even get converged results at all.\cite{ZenGeiRuaUmrCho-PRB-09}
For example, the local-density approximation (LDA) wrongly predicts largely delocalized 
electronic densities in strongly-correlated quantum wires,\cite{AbePolXiaTos-EJPB-07} being unable to 
reproduce the expected $N$-electron-peak structure due to
charge localization.\cite{AbePolXiaTos-EJPB-07}

Recently, a novel way of constructing 
exchange-correlation functionals for KS DFT has been 
proposed,\cite{MalGor-PRL-12,MalMirCreReiGor-PRB-13} based on the exact 
strong-coupling limit of DFT, which was formulated a few years ago within the 
so-called strictly-correlated-electrons (SCE) formalism.\cite{SeiGorSav-PRA-07,GorSeiVig-PRL-09,GorSei-PCCP-10} 
The first applications on quasi-one-dimensional quantum wires\cite{MalGor-PRL-12,MalMirCreReiGor-PRB-13} have 
shown that the resulting exchange-correlation functional is able to qualitatively describe arbitrary 
correlation regimes without artificially breaking any symmetry.  This is achieved since 
the SCE functional is able to create barriers (or ``bumps'') in the corresponding Kohn-Sham 
potential, which are a known feature of the exact one.\cite{BuiBaeSni-PRA-89,HelTokRub-JCP-09} 
Precisely these barriers can localize the charge density, avoiding the need of symmetry breaking 
to describe systems in which charge localization effects are important.
   
In this paper, we extend this approach to the study of two-dimensional
circularly-symmetric quantum dots with parabolic confinement. By considering different
confinement strengths, we investigate the crossover between the weakly-interacting 
and the strongly-correlated regimes. In particular, we reproduce well-known 
features of low-density quantum dots such as the formation of sharp rings 
in the electronic density\cite{GhoGucUmrUllBar-NP-06,GhoGucUmrUllBar-PRB-07} or the 
flattening of the addition energy spectra.\cite{YanLan-PRL-99}
Due to the relatively low computational cost of our approach,  we thus provide here an alternative powerful tool to study these kinds of systems.



\section{Quantum Dot Model}
\label{sec:model}

We consider two-dimensional (2D) quantum dots with $N$ electrons, laterally confined by a parabolic potential and described by the Hamiltonian (see, e.g., Ref.~\onlinecite{ReiMan-RMP-02})
\begin{equation}
\label{eq_HQD}
\hat{H} = \sum_{i=1}^N \left(-\frac{\hbar^2}{2 {m^*}}\nabla_i^2+m^* \frac{\omega^2}{2} r_i^2\right) + \frac{e^2}{\epsilon}\sum_{i=1}^{N-1} \sum_{j=i+1}^N \frac{1}{|\rv_i-\rv_j|},
\end{equation}
where $m^*$ is the effective mass and $\epsilon$ the dielectric constant. We use effective Hartree (H$^*$) units ($\hbar=1$, $a_B^*=\frac{\epsilon}{m^*} a_B=1$, $e=1$, $m^*=1$) throughout the rest of the paper.

The correlation regime is determined by the confinement strength: small (large) values 
of $\omega$ correspond to low (high) densities, for which the Coulomb repulsion 
dominates over (is dominated by) the kinetic energy. In order to characterize the
correlation quantitatively, one defines the so-called electron gas
parameter. In 2D it is given in terms of the electronic density 
as $r_s = (\pi\bar{n})^{-1/2}$, where $\bar{n}\equiv \int \rho(\rv)^2 d \rv/N$ is the average electron 
density. In the first calculations presented here, we have considered $N = 1, \dots, 10$ and $\omega\in[0.001,1]$, 
corresponding to values of $r_s$ between $\sim 1$ and 68. 


In the non-interacting case, the eigenfunctions of the system are the so-called Fock-Darwin
states,\cite{JacHawWoj-book-98} with associated energies given by
%
%
%
%
%
%
%
\begin{equation}
\label{FD_energies}
\varepsilon_{n,m} = 2\omega\left(n + \frac{1}{2} + \frac{|m|}{2}\right) \; ,
\end{equation}
where $m \in \Z$ and $n \in \N_0$ are, respectively, the angular and radial quantum 
numbers.


\section{Theoretical Approach}
\label{sec:method}

\subsection{KS DFT with the SCE functional}

We use the zeroth-order ``KS-SCE DFT'' approach, which was introduced in
Ref.~\onlinecite{MalGor-PRL-12} and described in more detail in Ref.~\onlinecite{MalMirCreReiGor-PRB-13}.
Essentially, the method consists of solving the standard spin-restricted Kohn-Sham 
equations,\cite{KohSha-PR-65}
\begin{equation}
\label{eq_KS}
\left(-\frac{\nabla^2}{2} + v_{\rm KS}[\rho](\rv) \right)\phi_i(\rv) 
= \varepsilon_i \phi_i(\rv) \; ,
\end{equation}
where $v_{\rm KS}[\rho](\rv)$ is the Kohn-Sham potential 
\begin{equation}
\label{eq_KS2}
v_{\rm KS}[\rho](\rv) = 
v_{\rm ext}[\rho](\rv) + v_{\rm Hartree}[\rho](\rv) + v_{\rm xc}[\rho](\rv) \; ,
\end{equation}
and $v_{\rm xc}[\rho](\rv)$ an approximate exchange-correlation potential that is constructed
from the functional derivative of the exact strong-interaction limit of the Hohenberg-Kohn functional. The resulting potential is able to capture the features of the strongly-correlated 
regime without introducing any spin or spatial symmetry breaking in the system. Below we 
briefly describe how the functional and the potential are built, and we refer the reader to 
Ref.~\onlinecite{MalMirCreReiGor-PRB-13} for further details.

The \emph{strictly-correlated-electrons} (SCE) functional $V_{\rm ee}^{\rm SCE}[\rho]$ of Seidl and 
co-workers\cite{Sei-PRA-99,SeiPerLev-PRA-99,SeiPerKur-PRL-00,SeiGorSav-PRA-07} is defined as the minimum possible electron-electron repulsion in \emph{a given smooth density} $\rho(\rv)$:
\begin{equation}
\label{eq_VeeSCE}
V_{\rm ee}^{\rm SCE}[\rho]\equiv \min_{\Psi\to\rho}\langle\Psi|\hat{V}_{\rm ee}|\Psi\rangle,
\end{equation}
where $\hat{V}_{\rm ee}$ is the Coulomb repulsion operator, i.e., the last term in Eq.~\eqref{eq_HQD}. It can be shown that in the low-density (or strong-interaction) limit, the Hohenberg-Kohn functional tends asymptotically to $V_{\rm ee}^{\rm SCE}[\rho]$.\cite{MalMirCreReiGor-PRB-13}
The SCE functional is the natural counterpart of the KS kinetic energy $T_s[\rho]$: the latter defines a reference system of non-interacting electrons with the same density of the physical system, while the former introduces a reference system (again with the same density) in which 
the electrons are infinitely (or perfectly) correlated, in the sense that the position of one of them determines all the relative positions 
in order to minimize the total Coulomb repulsion.

Thus, in the SCE system, if one electron (which we can label as ``1'' and take as a reference) is at 
position $\rv_1\equiv\rv$, the positions of the remaining $N-1$ electrons are given by the so-called 
\emph{co-motion functions}, $\rv_i\equiv \fv_i[\rho](\rv)$ ($i = 2, \dots, N$), which are non-local 
functionals of the density. They satisfy the differential equation\cite{SeiGorSav-PRA-07} 
\begin{equation}
\rho(\rv)d \rv = \rho(\fv_i(\rv)) d\fv_i(\rv)  \; ,
\label{eq_fi}
\end{equation}
or, equivalently, are such that the probability of finding one electron at 
position $\rv$ is the same of finding the electron $i$ at $\fv_i(\rv)$. The co-motion functions also satisfy the following cyclic group properties, which are a consequence of the indistinguishability of the electrons, ensuring that there is no dependence on which electron is chosen as reference, \cite{SeiGorSav-PRA-07}
\begin{equation}
\label{eq_groupprop}
\begin{split}
&\fv_1(\rv) \equiv \rv, \\
&\fv_2(\rv) \equiv \fv(\rv), \\
&\fv_3(\rv) =      \fv(\fv(\rv)), \\
&\fv_4(\rv) =      \fv(\fv(\fv(\rv))), \\
&\qquad\ \vdots \\
&\underbrace{\fv(\fv(\ldots\fv(\fv(\rv))))}_\text{$N$ times} = \rv.
\end{split}
\end{equation}
Notice that the SCE system describes a smooth $N$-electron quantum-mechanical 
density by means of an infinite superposition of degenerate classical 
configurations, which fulfill Eq.~\eqref{eq_fi} for every $\rv$. 
The square modulus of the corresponding wave function (which becomes a distribution in this limit\cite{ButDepGor-PRA-12,CotFriKlu-CPAM-13}) is given by 
\begin{multline}
\label{eq_psi2}
|\Psi_{\rm SCE}(\rv_1,\rv_2,\dots,\rv_N)|^2 = \frac{1}{N!} \sum_{\wp}
\int d\rv \, \frac{\rho(\rv)}{N} \, \delta(\rv_1-\fv_{\wp(1)}(\rv)) \\
\times\delta(\rv_2-\fv_{\wp(2)}(\rv)) \cdots \delta(\rv_N-\fv_{\wp(N)}(\rv))\; ,
\end{multline}
where $\wp$ denotes a permutation of ${1,\dots,N}$, such that 
$\rho(\rv) = N \int |\Psi_{\rm SCE}(\rv,\rv_2,\dots,\rv_N)|^2 
\,d\rv_2\cdots d\rv_N$.
The SCE system can thus be visualized as a ``floating'' Wigner crystal in a metric\cite{GorVigSei-JCTC-09} 
that describes the smooth density distribution $\rho(\rv)$.

In terms of the co-motion functions, the SCE functional $V_{\rm ee}^{\rm SCE}[\rho]$ 
of Eq.~\eqref{eq_VeeSCE} is given by\cite{SeiGorSav-PRA-07,MirSeiGor-JCTC-12}
\begin{equation}
\label{eq_VeeSCE2}
\begin{split}
V_{\rm ee}^{\rm SCE}[\rho]
&= \int d\rv\,\frac{\rho(\rv)}{N} \sum_{i=1}^{N-1}
\sum_{j=i+1}^N\frac{1}{|\fv_i(\rv)-\fv_j(\rv)|} \\
&= \frac{1}{2}\int d\rv\,\rho(\rv)\sum_{i= 2}^N \frac{1}{|\rv-\fv_i(\rv)|} \; .
\end{split}
\end{equation}
Another important property of the SCE system is the following: since the position 
of one electron at a given $\rv$ determines the other $N-1$ electronic
positions, the net Coulomb repulsion acting on an electron at a certain position $\rv$ 
becomes a function of $\rv$ itself. This force can be 
written in terms of the negative gradient of some one-body local 
potential $v_{\rm SCE}(\rv)$,\cite{MalMirCreReiGor-PRB-13} such that
\begin{equation}
\label{eq_vSCE}
-\nabla v_{\rm SCE}[\rho](\rv)
=\sum_{i=2}^N \frac{\rv-\fv_i[\rho](\rv)}{|\rv-\fv_i[\rho](\rv)|^3} \; .
\end{equation}
In turn, $v_{\rm SCE}[\rho](\rv)$ satisfies the important exact relation\cite{MalMirCreReiGor-PRB-13}
\begin{equation}
\label{eq_funcder}
v_{\rm SCE}[\rho](\rv) =
\frac{\delta V_{\rm ee}^{\rm SCE}[\rho]}{\delta \rho(\rv)} \; ,
\end{equation}
providing a very powerful shortcut for the construction of the functional derivative of $V_{\rm ee}^{\rm SCE}[\rho]$. 

The ``KS-SCE'' DFT approach to zeroth-order\cite{MalMirCreReiGor-PRB-13} 
consists in approximating the Hohenberg-Kohn functional as 
\begin{equation}
F[\rho] = T[\rho] + V_{\rm ee}[\rho]\simeq T_s[\rho] + V_{\rm ee}^{\rm SCE}[\rho] \; ,
\end{equation}
where $T_s[\rho]$ is the usual non-interacting Kohn-Sham kinetic energy.
By varying the total energy density functional
\begin{equation}
\label{eq_func}
E[\rho] \simeq T_s[\rho] + V_{\rm ee}^{\rm SCE}[\rho] + \int \rho(\rv) v_{\rm ext}(\rv)d\rv \;
\end{equation} 
with respect to the KS orbitals, and using Eq.~\eqref{eq_funcder}, we see that our approximation for the KS potential is 
\begin{equation}
v_{\rm KS}(\rv) \simeq v_{\rm ext}(\rv) + v_{\rm SCE}(\rv) \; 
\end{equation}
or, equivalently,
\begin{equation}
v_{\rm Hartree}(\rv) + v_{\rm xc}(\rv) \simeq v_{\rm SCE}(\rv) \; .
\end{equation}
Equation \eqref{eq_func} shows that the KS-SCE DFT approach treats both the kinetic 
energy and the electron-electron interaction on the same footing, letting the 
two terms compete in a self-consistent way within the Kohn-Sham scheme.
It can be shown that the method becomes asymptotically exact both in the very weak 
and very strong correlation limits.\cite{MalGor-PRL-12,MalMirCreReiGor-PRB-13} At 
intermediate correlation regimes it is expected to be less accurate,
but still qualitatively correct, as has already been shown 
when applied to one-dimensional quantum wires.\cite{MalMirCreReiGor-PRB-13}

\subsection{Practical implementation for circularly-symmetric 2D quantum dots}

The potential $v_{\rm SCE}(\rv)$ can be obtained by integrating Eq.~\eqref{eq_vSCE}. 
This requires the calculation of the co-motion functions $\fv_i(\rv)$ 
for a given density $\rho(\rv)$ via the solution of Eq.~\eqref{eq_fi}. 

For circularly symmetric two-dimensional systems, where the density depends 
only on the radial coordinate $r$, the problem can be separated into a radial 
and an angular part.\cite{SeiGorSav-PRA-07,GorSei-PCCP-10} The positions of the electrons, given by the co-motion 
functions, can then be expressed in polar coordinates 
as $\fv_i(\rv)=\fv_i(r,\theta)\equiv(f_i(r),\theta_i(r))$,
where the radial components satisfy Eq.~\eqref{eq_fi} rewritten
as \cite{SeiGorSav-PRA-07,GorSei-PCCP-10}
\begin{equation}
2\pi r \,\rho(r)\,dr = 2\pi f_i(r) \, \rho(f_i(r)) \, |f_i'(r)| \, dr \; .
\label{eq_fi_radial}
\end{equation}
These equations for the $f_i(r)$ can be solved by defining the function
\begin{equation}
N_e(r) = \int_0^r 2\pi r' \, \rho(r')\,dr' \;,
\label{eq_Ne}
\end{equation}
and its inverse $N_e^{-1}$.
The radial coordinates of the co-motion functions 
are then given by\cite{GorSei-PCCP-10}
\begin{equation}
\label{eq_fradial}
\begin{split}
f_{2k}(r) &= \begin{cases}
N_e^{-1}(2k-N_e(r)), & r \leq a_{2k} \\
N_e^{-1}(N_e(r)-2k), & r >    a_{2k}
\end{cases} \\
f_{2k+1}(r) &= \begin{cases}
N_e^{-1}(N_e(r)+2k),    & r\leq a_{N-2k} \\
N_e^{-1}(2N-2k-N_e(r)), & r >   a_{N-2k},
\end{cases}
\end{split}
\end{equation}
where  $a_k = N_e^{-1}(k)$, and the integer index $k$ runs from
1 to $(N-1)/2$ for odd $N$, and from 1 to $(N-2)/2$ for even $N$.
In the latter case, the $N^{\rm th}$ co-motion function is obtained
separately via
\begin{equation}
f_N(r) = N_e^{-1}(N-N_e(r)) \; .
\label{eq_fNeven}
\end{equation}
Equations \eqref{eq_Ne}--\eqref{eq_fNeven} show explicitly the non-local dependence of the $f_i(r)$ on $\rho(r)$.
One must then calculate the angular coordinates $\theta_i(r)$ of the co-motion 
functions as a function of $r$, the distance of one of the electrons from the center. These are obtained,\cite{SeiGorSav-PRA-07,GorSei-PCCP-10} for each value of $r$, by minimizing the total electron-electron repulsion energy
\begin{equation}
E_{\rm ee}(r) = \sum_{i>j} \Big(f_i(r)^{2} + f_j(r)^{2} - 
2f_i(r)f_j(r) \cos\theta_{ij}\Big)^{-1/2} \; ,
\end{equation}
%
with respect to the relative angles $\theta_{ij}=\theta_j-\theta_i$ between 
electrons $i$ and $j$ at positions $(f_i(r),\theta_i)$ and $(f_j(r),\theta_j)$. 

In two-dimensional problems, the number of relative angles to 
minimize is equal to $N-1$. In this first pilot implementation, whose primary goal is to check whether the KS SCE method is able to correctly describe the physics of low-density quantum dots, this angular minimization is done at each radial grid point, numerically. Thus, in each cycle of the self-consistent KS problem we perform $N_{\rm grid}$ times a $(N-1)$-dimensional minimization, where $N_{\rm grid}$ is  the number of grid points for the radial problem. As the angular minimization has also local minima, we proceed in the following way. For an initial non-degenerate radial configuration and given initial starting angles, we use the quasi-Newton Broyden-Fletcher-Goldfarb-Shanno (BFGS) algorithm to find the closest local minimum. Then we change the radial position of the ``first'' electron in small discrete steps, calculate the radial positions of the remaining electrons via Eqs.~\eqref{eq_fradial}-\eqref{eq_fNeven} and repeatedly optimize the angles using the BFGS algorithm, with starting angles taken from the previous step. This procedure rests on the assumption that the optimal angles change continuously with the radial configuration. Our numerical calculations suggest that this assumption is reasonable. Of course, the remaining open question is how to choose the starting angles for the initial radial configuration. We have experimented with simulated annealing as global optimization strategy. However, in these pilot applications we found it more practical to choose $N - 1$ pairwise different numbers ``by hand'' and probe several permutations of these numbers as starting angles. 

This strategy is by far not optimal, leaving space to several improvements that will be the object of future work. First of all, it should be noticed that the set of $N$ radial distances is periodic, as each circular shell $r\in[a_i, a_{i+1}]$ (with $a_i=N_e^{-1}(i)$, $i\in \N$), corresponds to the same physical situation,\cite{SeiGorSav-PRA-07} simply describing a permutation of the set of distances occurring in the first shell $r\in [0,a_1]$. Thus, by keeping track of the minimizing angles, and by readapting the grid in every circular shell, it is possible to do the angular minimization only $N_{\rm grid}/N$ times rather than $N_{\rm grid}$ times in each self-consistent field-iteration. 
Another important point that needs to be further investigated is the actual sensitivity of the results to the accuracy of the angular minimization. The optimal angles 
are used to determine the SCE potential by integrating Eq.~\eqref{eq_vSCE}, and we observe that this potential is not so sensitive to little variations of the optimal values, although a systematic study needs to be carried out. 

Overall, while in one dimension the SCE functional has a computational cost similar to LDA, in two dimensions the SCE is more expensive, because of the angular minimization. Still, its computational cost is much lower than that of wavefunction methods. Evaluating the total electron-electron repulsion energy scales like $\mathcal{O}(N^2)$ since all pairs of electrons have to be taken into account. The number of grid points can be made almost $N$-independent if we exploit the periodicity of the co-motion functions, so that one can always treat only one radial shell. With the local quasi-Newton scheme described above, we expect that the number of optimization steps increases moderately with $N$, such that the time complexity of our algorithm scales polynomially in $N$, with an exponent depending on how accurate the angular minimization needs actually to be. The storage requirements are also quite low compared to other methods like coupled cluster: storing all polar coordinates for the co-motion functions requires $\mathcal{O}(N_{\mathrm{grid}} \times N) = \mathcal{O}(N^2)$, which can be made linear in $N$ if we exploit the periodicity of the SCE problem. In future work, we will focus on optimizing the algorithm, studying larger numbers of particles.


\section{Results}
\label{sec:results}

\subsection{One-electron densities}

\begin{figure}
\includegraphics[width=0.85\columnwidth]{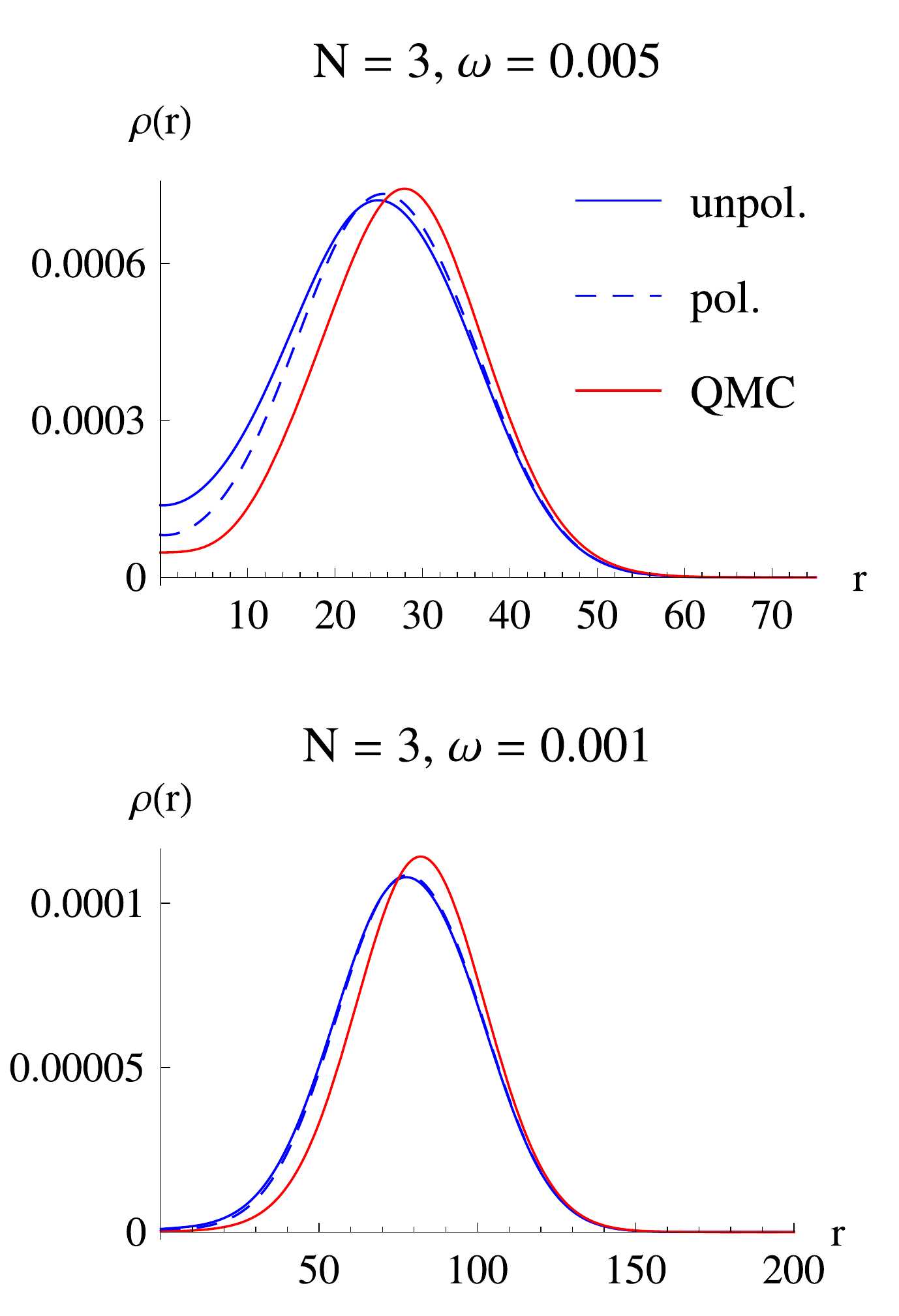}
\caption{(Color online) Comparison between the densities corresponding to $N=3$ 
obtained with the KS-SCE approach, both spin-polarized (dashed blue line) and 
spin-unpolarized (solid blue line) and with spin-polarized Quantum Monte Carlo (red line)
from Refs.~\onlinecite{GhoGucUmrUllBar-NP-06,GucGhoUmrBar-PRB-08}.}
\label{fig1}
\end{figure}

We have solved the self-consistent Kohn-Sham equations~\eqref{eq_KS} with the SCE potential for different values of the particle number $N$ and the confinement strength $\omega$. 

As mentioned, the main objective of this work is to show that KS DFT with the SCE functional is able to capture the features of the strongly-correlated regime without breaking any symmetry. A systematic comparison of the KS-SCE accuracy with available wavefunction results, as well as the optimization of the algorithm, will be the object of future work, where higher-order corrections to the SCE functional will also be developed and tested. Nonetheless, we want to provide an impression for the kind of quantitative accuracy that can be expected from our results. We thus compare, in Fig.~\ref{fig1}, the Quantum Monte Carlo densities of Refs.~\onlinecite{GhoGucUmrUllBar-NP-06,GucGhoUmrBar-PRB-08} for three-electron fully-spin-polarized 
quantum dots in the strongly-correlated regime with those obtained with our 
approach (both fully- and non-spin-polarized).
 It can be seen that already 
for $\omega=0.005$ the qualitative agreement is rather good, and that there 
is a small difference between the spin-polarized and unpolarized KS-SCE 
densities. As the correlation increases with smaller $\omega=0.001$, this difference 
becomes almost negligible as one would expect, and the agreement between our 
results and QMC improves.
It should be mentioned, however, that in contrast to the QMC calculations at these 
densities, the KS-SCE energy for the unpolarized cases has slightly lower energy than the 
spin-polarized solution. 
We attribute this discrepancy to the fact that the SCE functional, being intrinsically of 
classical nature,  is spin-independent and therefore unable 
to yield the lowest energy by occupying three different KS orbitals with the 
same spin. In future works we plan to add magnetic exchange and superexchange corrections to the SCE functional, which should allow the method to recognize the fully-spin-polarized solution as the ground-state one.
Quantitatively, the KS-SCE total energy has an error, with respect to QMC, of about 6 mH$^*$ ($\sim 6\%$) at $\omega=0.005$ and of about 1 mH$^*$ ($\sim 4\%$) at $\omega=0.001$. Notice that, while the fixed-node diffusion Monte Carlo provides an upper bound to the ground-state energy, the KS-SCE self-consistent energies are always a rigorous lower bound to the exact ground-state energy.\cite{MalGor-PRL-12,MalMirCreReiGor-PRB-13}

\begin{figure}
\includegraphics[width=\columnwidth]{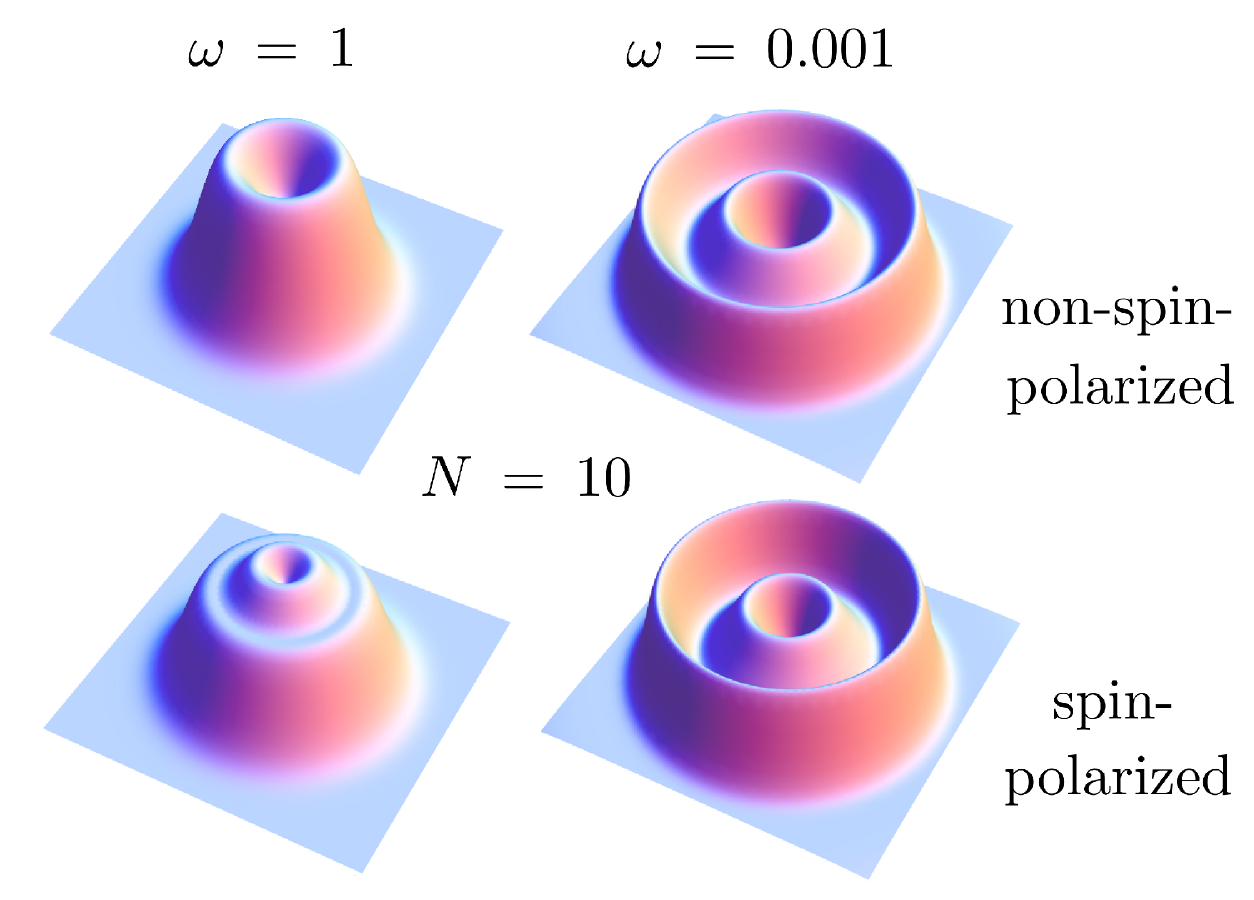}
\caption{(Color online) Electronic density $\rho(r)$ for a quantum dot with 
$N=10$ electrons, for $\omega=1$ and $\omega=0.001$.}
\label{fig2}
\end{figure}
We now illustrate and discuss the physical features of our results for the electronic densities, going from weakly to strongly-correlated quantum dots. Fig.~\ref{fig2} shows the self-consistent KS-SCE densities for quantum dots with $N=10$ electrons, 
considering both a strong ($\omega=1$) and a weak ($\omega=0.001$) confinement 
strength, for the fully-spin-polarized (one electron per Kohn-Sham orbital) 
and the non-spin-polarized (two electrons per orbital) cases. As mentioned above, when the confinement is strong the quantum 
dot is in the high-density regime and well described by the fermi-liquid 
shell structure, with a density distribution qualitatively similar to that 
obtained from the non-interacting Fock-Darwin states of Eq.~\eqref{FD_energies}.
Besides some slight oscillations due to the nodal structure of the different
orbitals, the resulting densities are rather ``thick'' or smoothed out, and, 
particularly in the spin-polarized dot, quite delocalized within the system. 
In both cases the values of the electron-gas parameter $r_s$ are $\simeq 1$.

As the confinement strength becomes weaker, the electron-electron
correlation plays an increasingly prominent role. The value $\omega=0.001$
corresponds to extremely low-density quantum dots, with $r_s\simeq 68$,
significantly larger than the maximum values achieved in previous works 
using wavefunction methods ($r_s\simeq 55$). \cite{GucGhoUmrBar-PRB-08} 
From the figure one can see how the density becomes much sharper in the 
radial direction, forming two very thin concentric rings centered at the 
origin. Integration of the density reveals the presence of two electrons 
in the inner ring and of 8 electrons in the outer one, in agreement with 
the ``8+2'' picture of the corresponding classical configuration made up 
of point-like charges --- see table 1 of Ref.~\onlinecite{BedPee-PRB-94}. 

It should be stressed that, as clearly seen from Fig.~\ref{fig2}, the densities 
obtained with the KS-SCE approach correctly\cite{GhoGucUmrUllBar-PRB-07} preserve the fundamental circular symmetry 
of the Hamiltonian of Eq.~\eqref{eq_HQD}. When the $v_{\rm SCE}(\rv)$ potential, which 
is constructed from the co-motion functions, is imported into the Kohn-Sham 
approach, it is able to describe properly the strongly-correlated regime, 
without introducing any artificial spatial or spin symmetry breaking. 
\begin{figure}
\includegraphics[width=7.0cm]{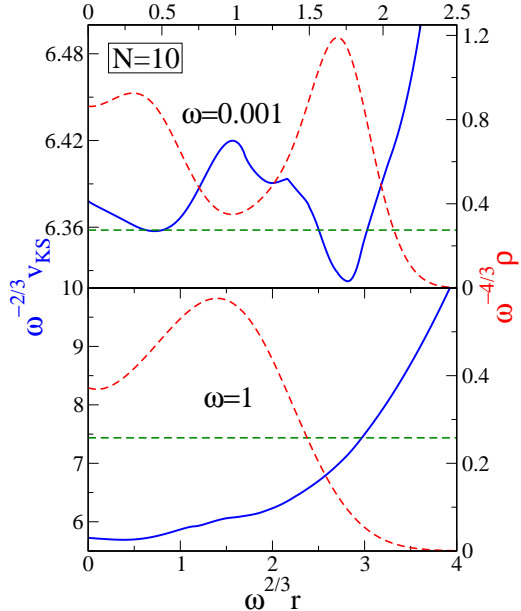}
\caption{(Color online) Self-consistent Kohn-Sham potentials (blue solid 
line) and densities (red dashed line) for the strongly- and 
weakly-correlated unpolarized cases of Fig.~\ref{fig2} (top and bottom, 
respectively). The green dashed horizontal lines correspond to the energies of the highest occupied KS 
orbitals. Notice the presence of classically forbidden regions inside the trap in the strongly-correlated case ($\omega=0.001$). }
\label{fig3}
\end{figure}
This happens because the SCE exchange-correlation potential 
self-consistently builds ``bumps'' that separate the charge density, capturing 
the physics of charge localization within the non-interacting KS formalism.
These structures were already observed in the case of one-dimensional quantum
wires using the KS-SCE approach,\cite{MalGor-PRL-12,MalMirCreReiGor-PRB-13}
with each maximum in the density corresponding to a minimum 
in the Kohn-Sham potential between consecutive ``bumps''. In Fig.~\ref{fig3} 
we show the self-consistent Kohn-Sham SCE potentials for 10-electron quantum dots with 
$\omega = 0.001$ and $\omega = 1$. Indeed, in the first case the potential
has a local maximum at the origin and a second one in the middle region,
giving rise to the density rings of Fig.~\ref{fig2}, also reported again in Fig.~\ref{fig3}.
In particular, the deep second minimum of the KS potential
is responsible for the sharp ring of the density in that region. In this way, restricted 
KS DFT reproduces the effect of strong correlation by means of a local one-body potential.
Conversely, in the weakly-interacting case $\omega=1$, the Kohn-Sham potential
does not display such structures. Here, the minimum of the density at the origin is not 
due to any maximum in the potential, but simply results from a fermionic-shell-structure 
effect. In the same Fig.~\ref{fig3} we also show, as horizontal green dashed lines, the highest occupied KS eigenvalue in both cases. One can clearly see
that in the strongly-correlated case ($\omega=0.001$) the barriers in the KS potential create classically forbidden regions inside the trap, giving rise to charge localization.

\begin{figure}
\includegraphics[width=7.5cm]{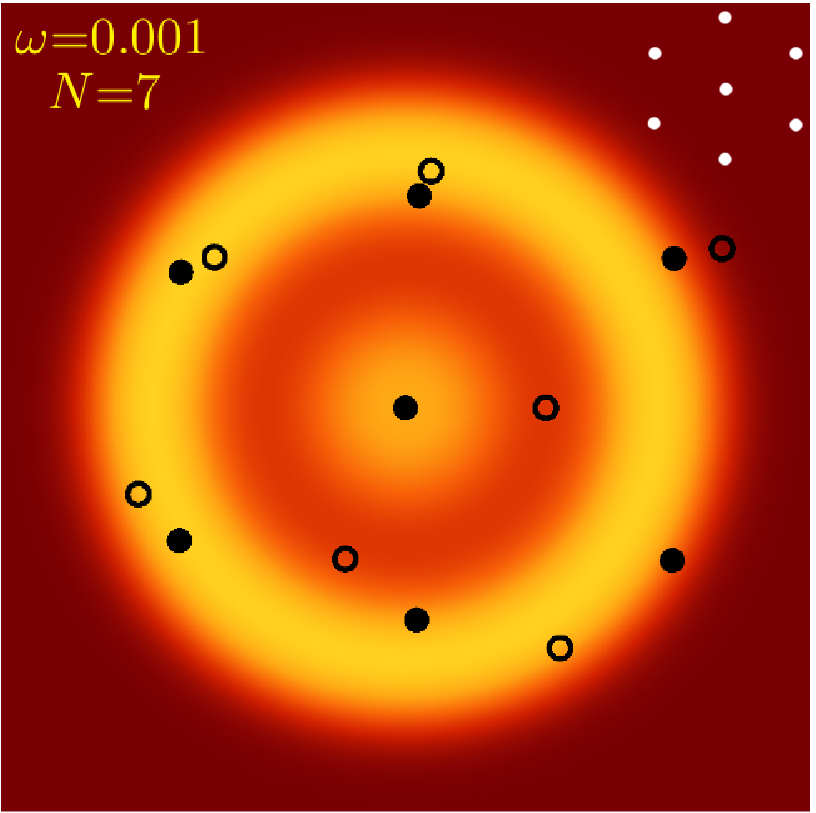}
\caption{(Color online) Co-motion functions $\fv_i(\rv)=(f_i(r),\theta_i(r))$ 
for two different configurations of the infinite superposition of
Eq.~\eqref{eq_psi2}, corresponding to the unpolarized dot with $N=7$ 
and $\omega=0.001$. Empty (solid) symbols represent the co-motion functions 
for a configuration with a small (large) weight. The classical ``1+6'' 
configuration is shown in the top right inset for the sake of comparison. The density is also shown as contour plot, with lighter colors indicating higher density regions.}
\label{fig4}
\end{figure}

In order to visualize the internal ordering of the electrons, in wavefunction methods 
one usually makes use of two-body quantities such as the pair-density 
distribution,\cite{GhoGucUmrUllBar-PRB-07} which is not accessible in density-functional 
approaches. 
Nevertheless, in the KS-SCE approach this internal ordering can be observed by looking at 
the co-motion functions of the SCE system, as we illustrate in Fig.~\ref{fig4}
for the unpolarized dot with $N=7$ and $\omega=0.001$. The figure shows the 
co-motion functions $\fv_i(\rv)=(f_i(r),\theta_i(r))$ corresponding to 
two different configurations, with large and small weight $\rho(r)/N$
in the infinite superposition of Eq.~\eqref{eq_psi2}, respectively. 
For this system, the density consists of a peak in the origin
(which integrates to one electron) and a sharp ring surrounding 
it (integrating to six electrons), as illustrated by the superimposed contour
plot (lighter colors: higher values of the density). The large-weight configuration is 
represented by solid symbols, and the low-weight one by empty symbols.
In the first case
the distribution of the co-motion functions closely resembles the 
classical point-charge configuration for this system, namely the 
``6+1'' distribution with one charge in the origin surrounded by 
an hexagon made up of the remaining six charges.\cite{BedPee-PRB-94} Notice that in order to yield 
a smooth density, also unusual configurations (like the one with empty symbols) need to have 
non-zero weight in the SCE $N$-body density of Eq.~\eqref{eq_psi2}. However, such configurations 
have a very small weight in the strongly-correlated regime.

\subsection{Addition energies}

\begin{figure}
\includegraphics[width=7.0cm]{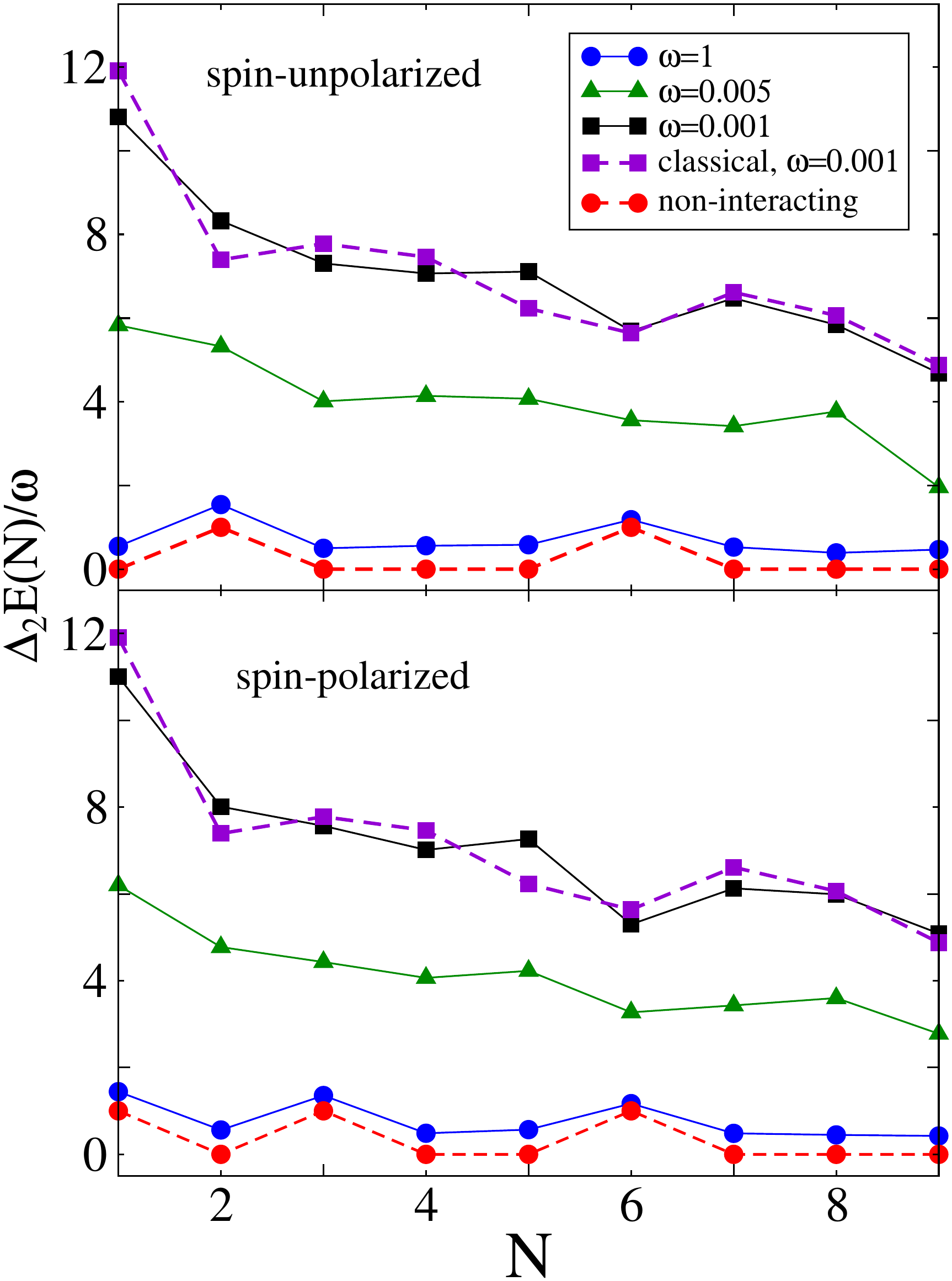}
\caption{(Color online) Addition energies as a function of $N$,
calculated via Eq.~\eqref{A2N} from the second difference in total energies,
for different values of the confinement strength $\omega$. For comparison, the noninteracting and classical cases are also shown.}
\label{fig5}
\end{figure}

\begin{figure}
\includegraphics[width=7.0cm]{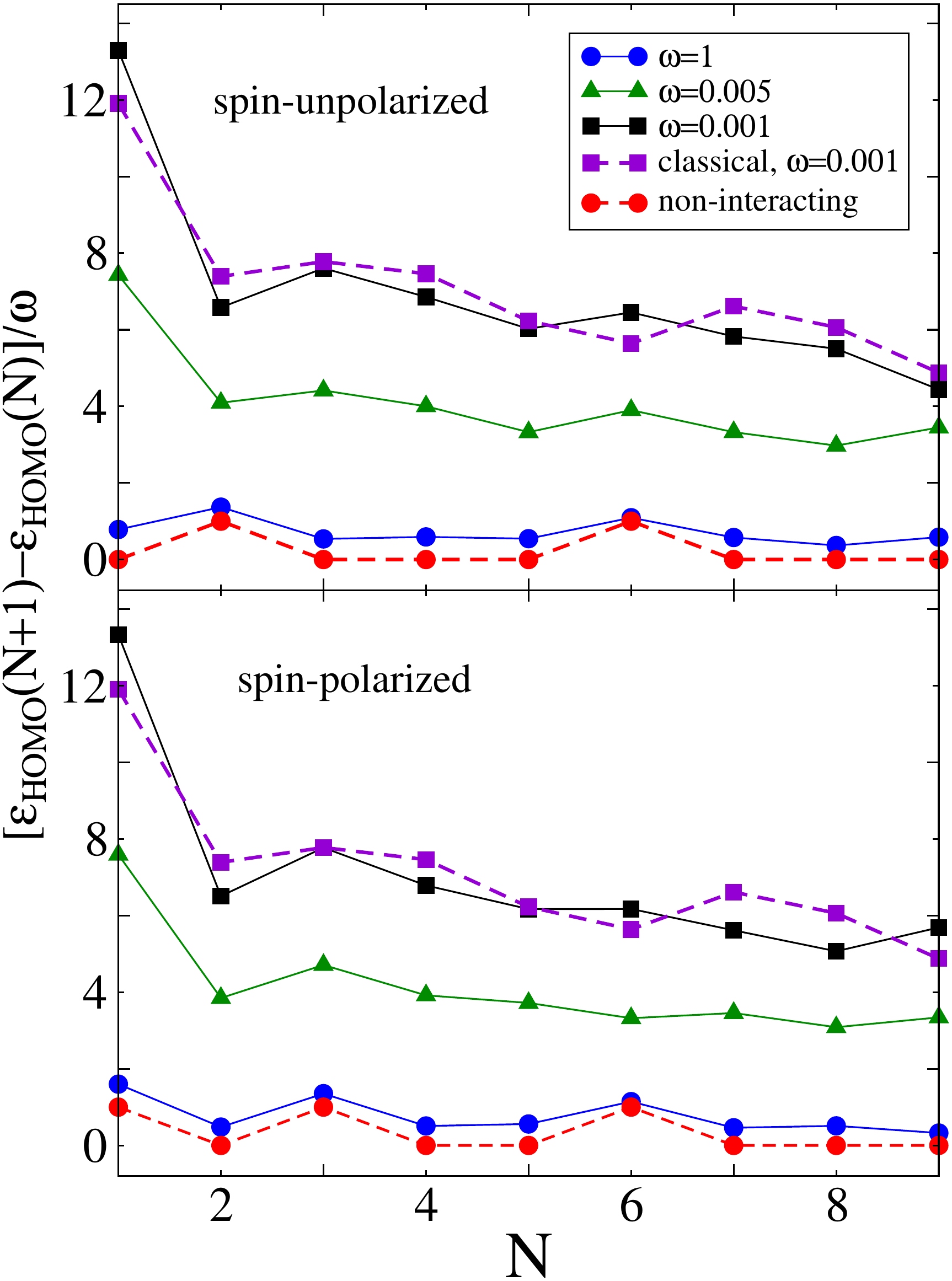}
\caption{(Color online) Same as Fig.~\ref{fig5}, but calculating the addition energies via Eq.~\eqref{Eadd} from the 
difference between the Kohn-Sham eigenvalues for the highest occupied orbitals.}
\label{fig6}
\end{figure}

In quantum-dot systems, the so-called addition energies provide useful information 
about the electronic structure of the system and can be probed 
experimentally.\cite{Ash-N-96,KouAusTar-RPP-01} 
They are defined as the second energy difference
\begin{equation}
\label{A2N}
E_{\rm add,E} \equiv \Delta_2 E(N) = E(N+1) - 2E(N) + E(N-1) \; ,
\end{equation}
where $E(N)$ is the total energy for the $N$-electron quantum dot. For KS DFT calculations, one can also use the alternative
expression
\begin{equation}
\label{Eadd}
E_{\rm add,HO}=\varepsilon_{\rm HO}(N+1)-\varepsilon_{\rm HO}(N) \; .
\end{equation}
It results from the fact that in the
exact KS theory --- that is, if the exact exchange-correlation potential 
were used --- the highest occupied (HO) Kohn-Sham eigenvalue $\varepsilon_{\rm HO}(N)$  is equal to minus the ionization energy of the physical, interacting, $N$-electron system,\cite{AlmBar-PRB-85,LevPerSah-PRA-84}
i.e., $\varepsilon_{\rm HO}(N) = E(N) - E(N-1)$. Notice that whereas the calculation of the addition energies using Eq.~\eqref{A2N}
requires knowledge about three different systems, the second alternative formula of
Eq.~\eqref{Eadd} only involves two of them. When using approximate functionals, the two expressions will not, in general, give the same results.

Figures \ref{fig5} and \ref{fig6} show the KS-SCE addition energies for quantum dots with up to $10$ electrons computed via Eq.~\eqref{A2N}
and Eq.~\eqref{Eadd}, respectively, for different strengths of the confining potential.
From both figures one can see that for strong confinement ($\omega=1$) the addition
energies are qualitatively similar to the non-interacting ones. In particular, 
for the non-spin-polarized systems they show the well-known peaks at $N=2$ and 6, 
corresponding to the closure of the first ($m=0$) and second ($m=\pm 1$) shells, 
and also the smaller peak at $N=4$ due to Hund's rule. In the spin-polarized
case, instead, the peaks are found at $N=1$ (first shell, $m=0$), at $N=3$ (second shell, $m = \pm 1$)
and at $N=5$ (third shell, $m = \pm 2$). 
When the quantum dots become strongly correlated, the shell structure changes radically.
The first main well-known feature is a flattening of the addition spectrum (notice that
in the figure the energies are divided by $\omega$, which in the low-density cases
takes values as low as $0.005$ and $0.001$). Secondly, the peak sequence becomes more irregular 
and resembles qualitatively the equivalent classical point-charge 
system.\cite{GucGhoUmrBar-PRB-08}


\section{Conclusions and Perspectives}
\label{sec:conclusions}

We have demonstrated the feasibility of constructing an exchange-correlation 
potential for spin-restricted Kohn-Sham density functional theory which is able to describe 
strong correlation effects in two-dimensional model quantum dots. This functional is derived
from the exact properties of the strong-coupling limit of the Hohenberg-Kohn functional.
It allows us to treat low-density quantum dots at relatively 
low computational cost when compared to other commonly employed approaches for 
studying these systems. Notice that, already for the number of particles and at the low densities considered here, CI calculations are not feasible. In the case of QMC, one has needed, so far, to make use of orbitals localized on different sites, thus breaking the circular symmetry of the system.\cite{GucGhoUmrBar-PRB-08} Our approach is numerically much less expensive, providing access to a broader parameter range than before. It also yields a set of radically new KS orbitals, which could be used in QMC instead of the localized gaussian ones. In other words, it would be very interesting to see if the KS-SCE orbitals provide good nodes for fixed-node diffusion Monte Carlo at low densities, avoiding the need of breaking the circular symmetry.\cite{ReiTouAssUmrHog-ACSSS-12} 

Overall, this new methodology shows the promise of becoming a powerful tool in low-dimensional, low-density, electronic structure calculations.
To exploit its full potential, several issues still need to be addressed in future works. First of all, corrections need to be designed to take the effects of the spin state in the SCE functional into account, for example using approximate magnetic exchange and superexchange functionals. Secondly, an efficient algorithm to solve (exactly or in a reasonably approximate way) the SCE equations for general (non-circularly symmetric) geometry needs to be fully developed. A viable route for this seems to be the dual Kantorovich formulation of the SCE functional,\cite{ButDepGor-PRA-12} whose first pilot implementation\cite{MenLin-PRB-13} has given promising results.

\subsection*{Acknowledgments} 
We thank Cyrus Umrigar and Devrim Guclu for the QMC densities of the $N = 3$ quantum dots and for interesting discussions. C.B.M. acknowledges support from the DFG 
project FR~1275/3-1. P.G-G. acknowledges support from the 
Netherlands Organization for Scientific Research (NWO) through a Vidi 
grant. Finally, this research was supported by a Marie Curie Intra European 
Fellowship within the 7th European Community Framework Programme (F.M.).


\begin{thebibliography}{51}
\expandafter\ifx\csname natexlab\endcsname\relax\def\natexlab#1{#1}\fi
\expandafter\ifx\csname bibnamefont\endcsname\relax
  \def\bibnamefont#1{#1}\fi
\expandafter\ifx\csname bibfnamefont\endcsname\relax
  \def\bibfnamefont#1{#1}\fi
\expandafter\ifx\csname citenamefont\endcsname\relax
  \def\citenamefont#1{#1}\fi
\expandafter\ifx\csname url\endcsname\relax
  \def\url#1{\texttt{#1}}\fi
\expandafter\ifx\csname urlprefix\endcsname\relax\def\urlprefix{URL }\fi
\providecommand{\bibinfo}[2]{#2}
\providecommand{\eprint}[2][]{\url{#2}}

\bibitem[{\citenamefont{Ghosal et~al.}(2006)\citenamefont{Ghosal, Guclu,
  Umrigar, Ullmo, and Baranger}}]{GhoGucUmrUllBar-NP-06}
\bibinfo{author}{\bibfnamefont{A.}~\bibnamefont{Ghosal}},
  \bibinfo{author}{\bibfnamefont{A.~D.} \bibnamefont{Guclu}},
  \bibinfo{author}{\bibfnamefont{C.~J.} \bibnamefont{Umrigar}},
  \bibinfo{author}{\bibfnamefont{D.}~\bibnamefont{Ullmo}}, \bibnamefont{and}
  \bibinfo{author}{\bibfnamefont{H.~U.} \bibnamefont{Baranger}},
  \bibinfo{journal}{Nature Phys.} \textbf{\bibinfo{volume}{2}},
  \bibinfo{pages}{336} (\bibinfo{year}{2006}).

\bibitem[{\citenamefont{Guclu et~al.}(2008)\citenamefont{Guclu, Ghosal,
  Umrigar, and Baranger}}]{GucGhoUmrBar-PRB-08}
\bibinfo{author}{\bibfnamefont{A.~D.} \bibnamefont{Guclu}},
  \bibinfo{author}{\bibfnamefont{A.}~\bibnamefont{Ghosal}},
  \bibinfo{author}{\bibfnamefont{C.~J.} \bibnamefont{Umrigar}},
  \bibnamefont{and} \bibinfo{author}{\bibfnamefont{H.~U.}
  \bibnamefont{Baranger}}, \bibinfo{journal}{Phys. Rev. B}
  \textbf{\bibinfo{volume}{77}}, \bibinfo{pages}{041301}
  (\bibinfo{year}{2008}).

\bibitem[{\citenamefont{Yannouleas and Landman}(2007)}]{YanLan-RPP-07}
\bibinfo{author}{\bibfnamefont{C.}~\bibnamefont{Yannouleas}} \bibnamefont{and}
  \bibinfo{author}{\bibfnamefont{U.}~\bibnamefont{Landman}},
  \bibinfo{journal}{Rep. Prog. Phys.} \textbf{\bibinfo{volume}{{70}}},
  \bibinfo{pages}{2067} (\bibinfo{year}{2007}).

\bibitem[{\citenamefont{Auslaender et~al.}(2005)\citenamefont{Auslaender,
  Steinberg, Yacoby, Tserkovnyak, Halperin, Baldwin, Pfeiffer, and
  West}}]{AusSteYacTseHalBalPfeWes-Sci-05}
\bibinfo{author}{\bibfnamefont{O.~M.} \bibnamefont{Auslaender}},
  \bibinfo{author}{\bibfnamefont{H.}~\bibnamefont{Steinberg}},
  \bibinfo{author}{\bibfnamefont{A.}~\bibnamefont{Yacoby}},
  \bibinfo{author}{\bibfnamefont{Y.}~\bibnamefont{Tserkovnyak}},
  \bibinfo{author}{\bibfnamefont{B.~I.} \bibnamefont{Halperin}},
  \bibinfo{author}{\bibfnamefont{K.~W.} \bibnamefont{Baldwin}},
  \bibinfo{author}{\bibfnamefont{L.~N.} \bibnamefont{Pfeiffer}},
  \bibnamefont{and} \bibinfo{author}{\bibfnamefont{K.~W.} \bibnamefont{West}},
  \bibinfo{journal}{Science (New York, N.Y.)} \textbf{\bibinfo{volume}{{308}}},
  \bibinfo{pages}{88} (\bibinfo{year}{2005}).

\bibitem[{\citenamefont{Kristinsdottir
  et~al.}(2011)\citenamefont{Kristinsdottir, Cremon, Nilsson, Xu, Samuelson,
  Linke, Wacker, and Reimann}}]{KriCreNilXuSamLinWacRei-PRB-11}
\bibinfo{author}{\bibfnamefont{L.~H.} \bibnamefont{Kristinsdottir}},
  \bibinfo{author}{\bibfnamefont{J.~C.} \bibnamefont{Cremon}},
  \bibinfo{author}{\bibfnamefont{H.~A.} \bibnamefont{Nilsson}},
  \bibinfo{author}{\bibfnamefont{H.~Q.} \bibnamefont{Xu}},
  \bibinfo{author}{\bibfnamefont{L.}~\bibnamefont{Samuelson}},
  \bibinfo{author}{\bibfnamefont{H.}~\bibnamefont{Linke}},
  \bibinfo{author}{\bibfnamefont{A.}~\bibnamefont{Wacker}}, \bibnamefont{and}
  \bibinfo{author}{\bibfnamefont{S.~M.} \bibnamefont{Reimann}},
  \bibinfo{journal}{Phys. Rev. B} \textbf{\bibinfo{volume}{{83}}},
  \bibinfo{pages}{041101} (\bibinfo{year}{2011}).

\bibitem[{\citenamefont{Creffield et~al.}(1999)\citenamefont{Creffield,
  Hausler, Jefferson, and Sarkar}}]{CreHauJefSar-PRB-99}
\bibinfo{author}{\bibfnamefont{C.~E.} \bibnamefont{Creffield}},
  \bibinfo{author}{\bibfnamefont{W.}~\bibnamefont{Hausler}},
  \bibinfo{author}{\bibfnamefont{J.~H.} \bibnamefont{Jefferson}},
  \bibnamefont{and} \bibinfo{author}{\bibfnamefont{S.}~\bibnamefont{Sarkar}},
  \bibinfo{journal}{Phys. Rev. B} \textbf{\bibinfo{volume}{{59}}},
  \bibinfo{pages}{10719} (\bibinfo{year}{1999}).

\bibitem[{\citenamefont{Egger et~al.}(1999)\citenamefont{Egger, Hausler, Mak,
  and Grabert}}]{EggHauMakGra-PRL-99}
\bibinfo{author}{\bibfnamefont{R.}~\bibnamefont{Egger}},
  \bibinfo{author}{\bibfnamefont{W.}~\bibnamefont{Hausler}},
  \bibinfo{author}{\bibfnamefont{C.~H.} \bibnamefont{Mak}}, \bibnamefont{and}
  \bibinfo{author}{\bibfnamefont{H.}~\bibnamefont{Grabert}},
  \bibinfo{journal}{Phys. Rev. Lett.} \textbf{\bibinfo{volume}{{82}}},
  \bibinfo{pages}{3320} (\bibinfo{year}{1999}).

\bibitem[{\citenamefont{Yannouleas and Landman}(1999)}]{YanLan-PRL-99}
\bibinfo{author}{\bibfnamefont{C.}~\bibnamefont{Yannouleas}} \bibnamefont{and}
  \bibinfo{author}{\bibfnamefont{U.}~\bibnamefont{Landman}},
  \bibinfo{journal}{Phys. Rev. Lett.} \textbf{\bibinfo{volume}{{82}}},
  \bibinfo{pages}{5325} (\bibinfo{year}{1999}).

\bibitem[{\citenamefont{Reimann et~al.}(2000)\citenamefont{Reimann, Koskinen,
  and Manninen}}]{ReiKosMan-PRB-00}
\bibinfo{author}{\bibfnamefont{S.}~\bibnamefont{Reimann}},
  \bibinfo{author}{\bibfnamefont{M.}~\bibnamefont{Koskinen}}, \bibnamefont{and}
  \bibinfo{author}{\bibfnamefont{M.}~\bibnamefont{Manninen}},
  \bibinfo{journal}{Phys. Rev. B} \textbf{\bibinfo{volume}{{62}}},
  \bibinfo{pages}{8108} (\bibinfo{year}{2000}).

\bibitem[{\citenamefont{Filinov et~al.}(2001)\citenamefont{Filinov, Bonitz, and
  Lozovik}}]{FilBonLoz-PRL-01}
\bibinfo{author}{\bibfnamefont{A.~V.} \bibnamefont{Filinov}},
  \bibinfo{author}{\bibfnamefont{M.}~\bibnamefont{Bonitz}}, \bibnamefont{and}
  \bibinfo{author}{\bibfnamefont{Y.~E.} \bibnamefont{Lozovik}},
  \bibinfo{journal}{Phys. Rev. Lett.} \textbf{\bibinfo{volume}{{86}}},
  \bibinfo{pages}{3851} (\bibinfo{year}{2001}).

\bibitem[{\citenamefont{Pederiva et~al.}(2002)\citenamefont{Pederiva,
  Emperador, and Lipparini}}]{PedEmpLip-PRB-02}
\bibinfo{author}{\bibfnamefont{F.}~\bibnamefont{Pederiva}},
  \bibinfo{author}{\bibfnamefont{A.}~\bibnamefont{Emperador}},
  \bibnamefont{and}
  \bibinfo{author}{\bibfnamefont{E.}~\bibnamefont{Lipparini}},
  \bibinfo{journal}{Phys. Rev. B} \textbf{\bibinfo{volume}{{66}}},
  \bibinfo{pages}{165314} (\bibinfo{year}{2002}).

\bibitem[{\citenamefont{Ghosal et~al.}(2007)\citenamefont{Ghosal, Guclu,
  Umrigar, Ullmo, and Baranger}}]{GhoGucUmrUllBar-PRB-07}
\bibinfo{author}{\bibfnamefont{A.}~\bibnamefont{Ghosal}},
  \bibinfo{author}{\bibfnamefont{A.~D.} \bibnamefont{Guclu}},
  \bibinfo{author}{\bibfnamefont{C.~J.} \bibnamefont{Umrigar}},
  \bibinfo{author}{\bibfnamefont{D.}~\bibnamefont{Ullmo}}, \bibnamefont{and}
  \bibinfo{author}{\bibfnamefont{H.~U.} \bibnamefont{Baranger}},
  \bibinfo{journal}{Phys. Rev. B} \textbf{\bibinfo{volume}{76}},
  \bibinfo{pages}{085341} (\bibinfo{year}{2007}).

\bibitem[{\citenamefont{Wigner}(1934)}]{Wig-PR-34}
\bibinfo{author}{\bibfnamefont{E.~P.} \bibnamefont{Wigner}},
  \bibinfo{journal}{Phys. Rev.} \textbf{\bibinfo{volume}{{46}}},
  \bibinfo{pages}{1002} (\bibinfo{year}{1934}).

\bibitem[{\citenamefont{Weiss et~al.}(2006)\citenamefont{Weiss, Thorwart, and
  Egger}}]{WeiThoEgg-EPL-06}
\bibinfo{author}{\bibfnamefont{S.}~\bibnamefont{Weiss}},
  \bibinfo{author}{\bibfnamefont{M.}~\bibnamefont{Thorwart}}, \bibnamefont{and}
  \bibinfo{author}{\bibfnamefont{R.}~\bibnamefont{Egger}},
  \bibinfo{journal}{Europhys. Lett.} \textbf{\bibinfo{volume}{{76}}},
  \bibinfo{pages}{905} (\bibinfo{year}{2006}).

\bibitem[{\citenamefont{Deshpande and Bockrath}(2008)}]{DesBoc-NP-08}
\bibinfo{author}{\bibfnamefont{V.~V.} \bibnamefont{Deshpande}}
  \bibnamefont{and} \bibinfo{author}{\bibfnamefont{M.}~\bibnamefont{Bockrath}},
  \bibinfo{journal}{Nature Phys.} \textbf{\bibinfo{volume}{{4}}},
  \bibinfo{pages}{314} (\bibinfo{year}{2008}).

\bibitem[{\citenamefont{Taylor and Calarco}(2008)}]{TayCal-PRA-08}
\bibinfo{author}{\bibfnamefont{J.~M.} \bibnamefont{Taylor}} \bibnamefont{and}
  \bibinfo{author}{\bibfnamefont{T.}~\bibnamefont{Calarco}},
  \bibinfo{journal}{Phys. Rev. A} \textbf{\bibinfo{volume}{{78}}},
  \bibinfo{pages}{062331} (\bibinfo{year}{2008}).

\bibitem[{\citenamefont{Ballester et~al.}(2010)\citenamefont{Ballester,
  Escart\'in, Movilla, Pi, and Planelles}}]{BalEscMovPiPla-PRB-10}
\bibinfo{author}{\bibfnamefont{A.}~\bibnamefont{Ballester}},
  \bibinfo{author}{\bibfnamefont{J.~M.} \bibnamefont{Escart\'in}},
  \bibinfo{author}{\bibfnamefont{J.~L.} \bibnamefont{Movilla}},
  \bibinfo{author}{\bibfnamefont{M.}~\bibnamefont{Pi}}, \bibnamefont{and}
  \bibinfo{author}{\bibfnamefont{J.}~\bibnamefont{Planelles}},
  \bibinfo{journal}{Phys. Rev. B} \textbf{\bibinfo{volume}{{82}}},
  \bibinfo{pages}{115405} (\bibinfo{year}{2010}).

\bibitem[{\citenamefont{Reimann and Manninen}(2002)}]{ReiMan-RMP-02}
\bibinfo{author}{\bibfnamefont{S.~M.} \bibnamefont{Reimann}} \bibnamefont{and}
  \bibinfo{author}{\bibfnamefont{M.}~\bibnamefont{Manninen}},
  \bibinfo{journal}{Rev. Mod. Phys.} \textbf{\bibinfo{volume}{{74}}},
  \bibinfo{pages}{1283} (\bibinfo{year}{2002}).

\bibitem[{\citenamefont{Rontani et~al.}(2006)\citenamefont{Rontani, Cavazzoni,
  Bellucci, and Goldoni}}]{RonCavBelGol-JCP-06}
\bibinfo{author}{\bibfnamefont{M.}~\bibnamefont{Rontani}},
  \bibinfo{author}{\bibfnamefont{C.}~\bibnamefont{Cavazzoni}},
  \bibinfo{author}{\bibfnamefont{D.}~\bibnamefont{Bellucci}}, \bibnamefont{and}
  \bibinfo{author}{\bibfnamefont{G.}~\bibnamefont{Goldoni}},
  \bibinfo{journal}{J. Chem. Phys.} \textbf{\bibinfo{volume}{{124}}},
  \bibinfo{pages}{124102} (\bibinfo{year}{2006}).

\bibitem[{\citenamefont{Waltersson et~al.}(2013)\citenamefont{Waltersson,
  Wessl\'en, and Lindroth}}]{WalWesLin-PRB-13}
\bibinfo{author}{\bibfnamefont{E.}~\bibnamefont{Waltersson}},
  \bibinfo{author}{\bibfnamefont{C.~J.} \bibnamefont{Wessl\'en}},
  \bibnamefont{and} \bibinfo{author}{\bibfnamefont{E.}~\bibnamefont{Lindroth}},
  \bibinfo{journal}{Phys. Rev. B} \textbf{\bibinfo{volume}{87}},
  \bibinfo{pages}{035112} (\bibinfo{year}{2013}).

\bibitem[{\citenamefont{Stoudenmire et~al.}(2012)\citenamefont{Stoudenmire,
  Wagner, White, and Burke}}]{StoWagWhiBur-PRL-12}
\bibinfo{author}{\bibfnamefont{E.}~\bibnamefont{Stoudenmire}},
  \bibinfo{author}{\bibfnamefont{L.~O.} \bibnamefont{Wagner}},
  \bibinfo{author}{\bibfnamefont{S.~R.} \bibnamefont{White}}, \bibnamefont{and}
  \bibinfo{author}{\bibfnamefont{K.}~\bibnamefont{Burke}},
  \bibinfo{journal}{Phys. Rev. Lett.} \textbf{\bibinfo{volume}{109}},
  \bibinfo{pages}{056402} (\bibinfo{year}{2012}).

\bibitem[{\citenamefont{Borgh et~al.}(2005)\citenamefont{Borgh, Toreblad,
  Koskinen, Manninen, Aberg, and Reimann}}]{BorTorKosManAbeRei-IJQC-05}
\bibinfo{author}{\bibfnamefont{M.}~\bibnamefont{Borgh}},
  \bibinfo{author}{\bibfnamefont{M.}~\bibnamefont{Toreblad}},
  \bibinfo{author}{\bibfnamefont{M.}~\bibnamefont{Koskinen}},
  \bibinfo{author}{\bibfnamefont{M.}~\bibnamefont{Manninen}},
  \bibinfo{author}{\bibfnamefont{S.}~\bibnamefont{Aberg}}, \bibnamefont{and}
  \bibinfo{author}{\bibfnamefont{S.~M.} \bibnamefont{Reimann}},
  \bibinfo{journal}{Int. J. Quantum Chem.} \textbf{\bibinfo{volume}{{105}}},
  \bibinfo{pages}{817} (\bibinfo{year}{2005}).

\bibitem[{\citenamefont{Anisimov et~al.}(1991)\citenamefont{Anisimov, Zaanen,
  and Andersen}}]{AniZaaAnd-PRB-91}
\bibinfo{author}{\bibfnamefont{V.~I.} \bibnamefont{Anisimov}},
  \bibinfo{author}{\bibfnamefont{J.}~\bibnamefont{Zaanen}}, \bibnamefont{and}
  \bibinfo{author}{\bibfnamefont{O.~K.} \bibnamefont{Andersen}},
  \bibinfo{journal}{Phys. Rev. B} \textbf{\bibinfo{volume}{44}},
  \bibinfo{pages}{943} (\bibinfo{year}{1991}).

\bibitem[{\citenamefont{Harju et~al.}(2004)\citenamefont{Harju, R\"as\"anen,
  Saarikoski, Puska, Nieminen, and Niemel\"a}}]{HarRasSaaPusNieNie-PRB-04}
\bibinfo{author}{\bibfnamefont{A.}~\bibnamefont{Harju}},
  \bibinfo{author}{\bibfnamefont{E.}~\bibnamefont{R\"as\"anen}},
  \bibinfo{author}{\bibfnamefont{H.}~\bibnamefont{Saarikoski}},
  \bibinfo{author}{\bibfnamefont{M.~J.} \bibnamefont{Puska}},
  \bibinfo{author}{\bibfnamefont{R.~M.} \bibnamefont{Nieminen}},
  \bibnamefont{and}
  \bibinfo{author}{\bibfnamefont{K.}~\bibnamefont{Niemel\"a}},
  \bibinfo{journal}{Phys. Rev. B} \textbf{\bibinfo{volume}{69}},
  \bibinfo{pages}{153101} (\bibinfo{year}{2004}).

\bibitem[{\citenamefont{Hohenberg and Kohn}(1964)}]{HohKoh-PR-64}
\bibinfo{author}{\bibfnamefont{P.}~\bibnamefont{Hohenberg}} \bibnamefont{and}
  \bibinfo{author}{\bibfnamefont{W.}~\bibnamefont{Kohn}},
  \bibinfo{journal}{Phys. Rev.} \textbf{\bibinfo{volume}{{136}}},
  \bibinfo{pages}{B 864} (\bibinfo{year}{1964}).

\bibitem[{\citenamefont{Kohn and Sham}(1965)}]{KohSha-PR-65}
\bibinfo{author}{\bibfnamefont{W.}~\bibnamefont{Kohn}} \bibnamefont{and}
  \bibinfo{author}{\bibfnamefont{L.~J.} \bibnamefont{Sham}},
  \bibinfo{journal}{Phys. Rev. A} \textbf{\bibinfo{volume}{140}},
  \bibinfo{pages}{1133} (\bibinfo{year}{1965}).

\bibitem[{\citenamefont{Cohen et~al.}(2008)\citenamefont{Cohen, Mori-Sanchez,
  and Yang}}]{CohMorYan-SCI-08}
\bibinfo{author}{\bibfnamefont{A.~J.} \bibnamefont{Cohen}},
  \bibinfo{author}{\bibfnamefont{P.}~\bibnamefont{Mori-Sanchez}},
  \bibnamefont{and} \bibinfo{author}{\bibfnamefont{W.~T.} \bibnamefont{Yang}},
  \bibinfo{journal}{Science} \textbf{\bibinfo{volume}{{321}}},
  \bibinfo{pages}{792} (\bibinfo{year}{2008}).

\bibitem[{\citenamefont{Abedinpour et~al.}(2007)\citenamefont{Abedinpour,
  Polini, Xianlong, and Tosi}}]{AbePolXiaTos-EJPB-07}
\bibinfo{author}{\bibfnamefont{S.~H.} \bibnamefont{Abedinpour}},
  \bibinfo{author}{\bibfnamefont{M.}~\bibnamefont{Polini}},
  \bibinfo{author}{\bibfnamefont{G.}~\bibnamefont{Xianlong}}, \bibnamefont{and}
  \bibinfo{author}{\bibfnamefont{M.~P.} \bibnamefont{Tosi}},
  \bibinfo{journal}{Eur. Phys. J. B} \textbf{\bibinfo{volume}{{56}}},
  \bibinfo{pages}{127} (\bibinfo{year}{2007}).

\bibitem[{\citenamefont{Zeng et~al.}(2009)\citenamefont{Zeng, Geist, Ruan,
  Umrigar, and Chou}}]{ZenGeiRuaUmrCho-PRB-09}
\bibinfo{author}{\bibfnamefont{L.}~\bibnamefont{Zeng}},
  \bibinfo{author}{\bibfnamefont{W.}~\bibnamefont{Geist}},
  \bibinfo{author}{\bibfnamefont{W.~Y.} \bibnamefont{Ruan}},
  \bibinfo{author}{\bibfnamefont{C.~J.} \bibnamefont{Umrigar}},
  \bibnamefont{and} \bibinfo{author}{\bibfnamefont{M.~Y.} \bibnamefont{Chou}},
  \bibinfo{journal}{Phys. Rev. B} \textbf{\bibinfo{volume}{79}},
  \bibinfo{pages}{235334} (\bibinfo{year}{2009}).

\bibitem[{\citenamefont{Malet and Gori-Giorgi}(2012)}]{MalGor-PRL-12}
\bibinfo{author}{\bibfnamefont{F.}~\bibnamefont{Malet}} \bibnamefont{and}
  \bibinfo{author}{\bibfnamefont{P.}~\bibnamefont{Gori-Giorgi}},
  \bibinfo{journal}{Phys. Rev. Lett.} \textbf{\bibinfo{volume}{{109}}},
  \bibinfo{pages}{246402} (\bibinfo{year}{2012}).

\bibitem[{\citenamefont{Malet et~al.}(2013)\citenamefont{Malet, Mirtschink,
  Cremon, Reimann, and Gori-Giorgi}}]{MalMirCreReiGor-PRB-13}
\bibinfo{author}{\bibfnamefont{F.}~\bibnamefont{Malet}},
  \bibinfo{author}{\bibfnamefont{A.}~\bibnamefont{Mirtschink}},
  \bibinfo{author}{\bibfnamefont{J.~C.} \bibnamefont{Cremon}},
  \bibinfo{author}{\bibfnamefont{S.~M.} \bibnamefont{Reimann}},
  \bibnamefont{and}
  \bibinfo{author}{\bibfnamefont{P.}~\bibnamefont{Gori-Giorgi}},
  \bibinfo{journal}{Phys. Rev. B} \textbf{\bibinfo{volume}{{87}}},
  \bibinfo{pages}{115146} (\bibinfo{year}{2013}).

\bibitem[{\citenamefont{Seidl et~al.}(2007)\citenamefont{Seidl, Gori-Giorgi,
  and Savin}}]{SeiGorSav-PRA-07}
\bibinfo{author}{\bibfnamefont{M.}~\bibnamefont{Seidl}},
  \bibinfo{author}{\bibfnamefont{P.}~\bibnamefont{Gori-Giorgi}},
  \bibnamefont{and} \bibinfo{author}{\bibfnamefont{A.}~\bibnamefont{Savin}},
  \bibinfo{journal}{Phys. Rev. A} \textbf{\bibinfo{volume}{{75}}},
  \bibinfo{pages}{042511} (\bibinfo{year}{2007}).

\bibitem[{\citenamefont{Gori-Giorgi
  et~al.}(2009{\natexlab{a}})\citenamefont{Gori-Giorgi, Seidl, and
  Vignale}}]{GorSeiVig-PRL-09}
\bibinfo{author}{\bibfnamefont{P.}~\bibnamefont{Gori-Giorgi}},
  \bibinfo{author}{\bibfnamefont{M.}~\bibnamefont{Seidl}}, \bibnamefont{and}
  \bibinfo{author}{\bibfnamefont{G.}~\bibnamefont{Vignale}},
  \bibinfo{journal}{Phys. Rev. Lett.} \textbf{\bibinfo{volume}{{103}}},
  \bibinfo{pages}{166402} (\bibinfo{year}{2009}{\natexlab{a}}).

\bibitem[{\citenamefont{Gori-Giorgi and Seidl}(2010)}]{GorSei-PCCP-10}
\bibinfo{author}{\bibfnamefont{P.}~\bibnamefont{Gori-Giorgi}} \bibnamefont{and}
  \bibinfo{author}{\bibfnamefont{M.}~\bibnamefont{Seidl}},
  \bibinfo{journal}{Phys. Chem. Chem. Phys.} \textbf{\bibinfo{volume}{{12}}},
  \bibinfo{pages}{14405} (\bibinfo{year}{2010}).

\bibitem[{\citenamefont{Buijse et~al.}(1989)\citenamefont{Buijse, Baerends, and
  Snijders}}]{BuiBaeSni-PRA-89}
\bibinfo{author}{\bibfnamefont{M.~A.} \bibnamefont{Buijse}},
  \bibinfo{author}{\bibfnamefont{E.~J.} \bibnamefont{Baerends}},
  \bibnamefont{and} \bibinfo{author}{\bibfnamefont{J.~G.}
  \bibnamefont{Snijders}}, \bibinfo{journal}{Phys. Rev. A}
  \textbf{\bibinfo{volume}{{40}}}, \bibinfo{pages}{4190}
  (\bibinfo{year}{1989}).

\bibitem[{\citenamefont{Helbig et~al.}(2009)\citenamefont{Helbig, Tokatly, and
  Rubio}}]{HelTokRub-JCP-09}
\bibinfo{author}{\bibfnamefont{N.}~\bibnamefont{Helbig}},
  \bibinfo{author}{\bibfnamefont{I.~V.} \bibnamefont{Tokatly}},
  \bibnamefont{and} \bibinfo{author}{\bibfnamefont{A.}~\bibnamefont{Rubio}},
  \bibinfo{journal}{J. Chem. Phys.} \textbf{\bibinfo{volume}{131}},
  \bibinfo{pages}{224105} (\bibinfo{year}{2009}).

\bibitem[{\citenamefont{Jacak et~al.}(1998)\citenamefont{Jacak, Hawrylak, and
  W\'ojs}}]{JacHawWoj-book-98}
\bibinfo{author}{\bibfnamefont{L.}~\bibnamefont{Jacak}},
  \bibinfo{author}{\bibfnamefont{P.}~\bibnamefont{Hawrylak}}, \bibnamefont{and}
  \bibinfo{author}{\bibfnamefont{A.}~\bibnamefont{W\'ojs}},
  \emph{\bibinfo{title}{Quantum Dots}} (\bibinfo{publisher}{Springer},
  \bibinfo{address}{Berlin}, \bibinfo{year}{1998}).

\bibitem[{\citenamefont{Seidl}(1999)}]{Sei-PRA-99}
\bibinfo{author}{\bibfnamefont{M.}~\bibnamefont{Seidl}},
  \bibinfo{journal}{Phys. Rev. A} \textbf{\bibinfo{volume}{{60}}},
  \bibinfo{pages}{4387} (\bibinfo{year}{1999}).

\bibitem[{\citenamefont{Seidl et~al.}(1999)\citenamefont{Seidl, Perdew, and
  Levy}}]{SeiPerLev-PRA-99}
\bibinfo{author}{\bibfnamefont{M.}~\bibnamefont{Seidl}},
  \bibinfo{author}{\bibfnamefont{J.~P.} \bibnamefont{Perdew}},
  \bibnamefont{and} \bibinfo{author}{\bibfnamefont{M.}~\bibnamefont{Levy}},
  \bibinfo{journal}{Phys. Rev. A} \textbf{\bibinfo{volume}{{59}}},
  \bibinfo{pages}{51} (\bibinfo{year}{1999}).

\bibitem[{\citenamefont{Seidl et~al.}(2000)\citenamefont{Seidl, Perdew, and
  Kurth}}]{SeiPerKur-PRL-00}
\bibinfo{author}{\bibfnamefont{M.}~\bibnamefont{Seidl}},
  \bibinfo{author}{\bibfnamefont{J.~P.} \bibnamefont{Perdew}},
  \bibnamefont{and} \bibinfo{author}{\bibfnamefont{S.}~\bibnamefont{Kurth}},
  \bibinfo{journal}{Phys. Rev. Lett.} \textbf{\bibinfo{volume}{{84}}},
  \bibinfo{pages}{5070} (\bibinfo{year}{2000}).

\bibitem[{\citenamefont{Buttazzo et~al.}(2012)\citenamefont{Buttazzo, {De
  Pascale}, and Gori-Giorgi}}]{ButDepGor-PRA-12}
\bibinfo{author}{\bibfnamefont{G.}~\bibnamefont{Buttazzo}},
  \bibinfo{author}{\bibfnamefont{L.}~\bibnamefont{{De Pascale}}},
  \bibnamefont{and}
  \bibinfo{author}{\bibfnamefont{P.}~\bibnamefont{Gori-Giorgi}},
  \bibinfo{journal}{Phys. Rev. A} \textbf{\bibinfo{volume}{{85}}},
  \bibinfo{pages}{062502} (\bibinfo{year}{2012}).

\bibitem[{\citenamefont{Cotar et~al.}(2013)\citenamefont{Cotar, Friesecke, and
  Kl\"uppelberg}}]{CotFriKlu-CPAM-13}
\bibinfo{author}{\bibfnamefont{C.}~\bibnamefont{Cotar}},
  \bibinfo{author}{\bibfnamefont{G.}~\bibnamefont{Friesecke}},
  \bibnamefont{and}
  \bibinfo{author}{\bibfnamefont{C.}~\bibnamefont{Kl\"uppelberg}},
  \bibinfo{journal}{Comm. Pure Appl. Math.} \textbf{\bibinfo{volume}{66}},
  \bibinfo{pages}{548} (\bibinfo{year}{2013}).

\bibitem[{\citenamefont{Gori-Giorgi
  et~al.}(2009{\natexlab{b}})\citenamefont{Gori-Giorgi, Vignale, and
  Seidl}}]{GorVigSei-JCTC-09}
\bibinfo{author}{\bibfnamefont{P.}~\bibnamefont{Gori-Giorgi}},
  \bibinfo{author}{\bibfnamefont{G.}~\bibnamefont{Vignale}}, \bibnamefont{and}
  \bibinfo{author}{\bibfnamefont{M.}~\bibnamefont{Seidl}}, \bibinfo{journal}{J.
  Chem. Theory Comput.} \textbf{\bibinfo{volume}{{5}}}, \bibinfo{pages}{743}
  (\bibinfo{year}{2009}{\natexlab{b}}).

\bibitem[{\citenamefont{Mirtschink et~al.}(2012)\citenamefont{Mirtschink,
  Seidl, and Gori-Giorgi}}]{MirSeiGor-JCTC-12}
\bibinfo{author}{\bibfnamefont{A.}~\bibnamefont{Mirtschink}},
  \bibinfo{author}{\bibfnamefont{M.}~\bibnamefont{Seidl}}, \bibnamefont{and}
  \bibinfo{author}{\bibfnamefont{P.}~\bibnamefont{Gori-Giorgi}},
  \bibinfo{journal}{J. Chem. Theory Comput.} \textbf{\bibinfo{volume}{8}},
  \bibinfo{pages}{3097} (\bibinfo{year}{2012}).

\bibitem[{\citenamefont{Bedanov and Peeters}(1994)}]{BedPee-PRB-94}
\bibinfo{author}{\bibfnamefont{V.~M.} \bibnamefont{Bedanov}} \bibnamefont{and}
  \bibinfo{author}{\bibfnamefont{F.~M.} \bibnamefont{Peeters}},
  \bibinfo{journal}{Phys. Rev. B} \textbf{\bibinfo{volume}{49}},
  \bibinfo{pages}{2667} (\bibinfo{year}{1994}).

\bibitem[{\citenamefont{Ashoori}(1996)}]{Ash-N-96}
\bibinfo{author}{\bibfnamefont{R.~C.} \bibnamefont{Ashoori}},
  \bibinfo{journal}{Nature} \textbf{\bibinfo{volume}{379}},
  \bibinfo{pages}{413} (\bibinfo{year}{1996}).

\bibitem[{\citenamefont{Kouwenhoven et~al.}(2001)\citenamefont{Kouwenhoven,
  Austing, and Tarucha}}]{KouAusTar-RPP-01}
\bibinfo{author}{\bibfnamefont{L.~P.} \bibnamefont{Kouwenhoven}},
  \bibinfo{author}{\bibfnamefont{D.~G.} \bibnamefont{Austing}},
  \bibnamefont{and} \bibinfo{author}{\bibfnamefont{S.}~\bibnamefont{Tarucha}},
  \bibinfo{journal}{Rep. Prog. Phys.} \textbf{\bibinfo{volume}{64}},
  \bibinfo{pages}{701} (\bibinfo{year}{2001}).

\bibitem[{\citenamefont{Almbladh and von Barth}(1985)}]{AlmBar-PRB-85}
\bibinfo{author}{\bibfnamefont{C.-O.} \bibnamefont{Almbladh}} \bibnamefont{and}
  \bibinfo{author}{\bibfnamefont{U.}~\bibnamefont{von Barth}},
  \bibinfo{journal}{Phys. Rev. B} \textbf{\bibinfo{volume}{31}},
  \bibinfo{pages}{3231} (\bibinfo{year}{1985}).

\bibitem[{\citenamefont{Levy et~al.}(1984)\citenamefont{Levy, Perdew, and
  Sahni}}]{LevPerSah-PRA-84}
\bibinfo{author}{\bibfnamefont{M.}~\bibnamefont{Levy}},
  \bibinfo{author}{\bibfnamefont{J.~P.} \bibnamefont{Perdew}},
  \bibnamefont{and} \bibinfo{author}{\bibfnamefont{V.}~\bibnamefont{Sahni}},
  \bibinfo{journal}{Phys. Rev. A} \textbf{\bibinfo{volume}{30}},
  \bibinfo{pages}{2745} (\bibinfo{year}{1984}).

\bibitem[{\citenamefont{Reinhardt et~al.}(2012)\citenamefont{Reinhardt,
  Toulouse, Assaraf, Umrigar, and Hoggan}}]{ReiTouAssUmrHog-ACSSS-12}
\bibinfo{author}{\bibfnamefont{P.}~\bibnamefont{Reinhardt}},
  \bibinfo{author}{\bibfnamefont{J.}~\bibnamefont{Toulouse}},
  \bibinfo{author}{\bibfnamefont{R.}~\bibnamefont{Assaraf}},
  \bibinfo{author}{\bibfnamefont{C.~J.} \bibnamefont{Umrigar}},
  \bibnamefont{and} \bibinfo{author}{\bibfnamefont{P.~E.}
  \bibnamefont{Hoggan}}, \bibinfo{journal}{ACS Symposium Series}
  \textbf{\bibinfo{volume}{{53}}}, \bibinfo{pages}{1094}
  (\bibinfo{year}{2012}).

\bibitem[{\citenamefont{Mendl and Lin}(2013)}]{MenLin-PRB-13}
\bibinfo{author}{\bibfnamefont{C.~B.} \bibnamefont{Mendl}} \bibnamefont{and}
  \bibinfo{author}{\bibfnamefont{L.}~\bibnamefont{Lin}},
  \bibinfo{journal}{Phys. Rev. B} \textbf{\bibinfo{volume}{87}},
  \bibinfo{pages}{125106} (\bibinfo{year}{2013}).

\end{thebibliography}

\end{document}